\documentclass[aps,prd,twocolumn,groupedaddress,superscriptaddress]{revtex4-2}

\usepackage{lipsum}
\usepackage{graphicx}  
\usepackage{dcolumn}   
\usepackage{bm}        
\usepackage{amssymb}   
\usepackage{amsmath}
\usepackage{xcolor}
\usepackage{hyperref}
\usepackage{amsmath}
\usepackage{amsfonts}

\hypersetup{colorlinks=true, linkcolor=blue, citecolor=blue, urlcolor=blue}

\begin{document}


\title{\bf Search for galactic axions with a high-Q dielectric cavity}

\author{D.~Alesini} \affiliation{INFN, Laboratori Nazionali di Frascati, Frascati, Roma, Italy}
\author{D.~Babusci} \affiliation{INFN, Laboratori Nazionali di Frascati, Frascati, Roma, Italy}
\author{C.~Braggio} \affiliation{INFN, Sezione di Padova, Padova, Italy} \affiliation{Dipartimento di Fisica e Astronomia, Padova, Italy}
\author{G.~Carugno} \affiliation{INFN, Sezione di Padova, Padova, Italy} \affiliation{Dipartimento di Fisica e Astronomia, Padova, Italy}
\author{N.~Crescini} \altaffiliation{Present address: University Grenoble Alpes, CNRS, Grenoble INP, Institut Néel, 38000 Grenoble, France} \affiliation{INFN, Laboratori Nazionali di Legnaro, Legnaro, Padova, Italy} \affiliation{Dipartimento di Fisica e Astronomia, Padova, Italy}
\author{D.~D'Agostino} \affiliation{Dipartimento di Fisica E.R. Caianiello, Fisciano, Salerno, Italy} \affiliation{INFN, Sezione di Napoli, Napoli, Italy}
\author{A.~D'Elia}\email{alessandro.delia@lnf.infn.it} \affiliation{INFN, Laboratori Nazionali di Frascati, Frascati, Roma, Italy}
\author{D.~Di~Gioacchino} \affiliation{INFN, Laboratori Nazionali di Frascati, Frascati, Roma, Italy}
\author{R.~Di Vora}\email{divora@pd.infn.it} \affiliation{INFN, Sezione di Padova, Padova, Italy} \affiliation{Dipartimento di Scienze Fisiche, della Terra e dell'Ambiente, Universit{\`a} di Siena,  53100 Siena, Italy}
\author{P.~Falferi} \affiliation{Istituto di Fotonica e Nanotecnologie, CNR Fondazione Bruno Kessler, I-38123 Povo, Trento, Italy} \affiliation{INFN, TIFPA, Povo, Trento, Italy}
\author{U.~Gambardella} \affiliation{Dipartimento di Fisica E.R. Caianiello, Fisciano, Salerno, Italy} \affiliation{INFN, Sezione di Napoli, Napoli, Italy}
\author{C.~Gatti} \affiliation{INFN, Laboratori Nazionali di Frascati, Frascati, Roma, Italy}
\author{G.~Iannone} \affiliation{Dipartimento di Fisica E.R. Caianiello, Fisciano, Salerno, Italy} \affiliation{INFN, Sezione di Napoli, Napoli, Italy}
\author{C.~Ligi} \affiliation{INFN, Laboratori Nazionali di Frascati, Frascati, Roma, Italy}
\author{A.~Lombardi} \affiliation{INFN, Laboratori Nazionali di Legnaro, Legnaro, Padova, Italy}
\author{G.~Maccarrone} \affiliation{INFN, Laboratori Nazionali di Frascati, Frascati, Roma, Italy}
\author{A.~Ortolan} \affiliation{INFN, Laboratori Nazionali di Legnaro, Legnaro, Padova, Italy}
\author{R.~Pengo} \affiliation{INFN, Laboratori Nazionali di Legnaro, Legnaro, Padova, Italy}
\author{A.~Rettaroli} \affiliation{INFN, Laboratori Nazionali di Frascati, Frascati, Roma, Italy} 
\author{G.~Ruoso} \affiliation{INFN, Laboratori Nazionali di Legnaro, Legnaro, Padova, Italy}
\author{L.~Taffarello} \affiliation{INFN, Sezione di Padova, Padova, Italy}
\author{S.~Tocci} \affiliation{INFN, Laboratori Nazionali di Frascati, Frascati, Roma, Italy}

\date{\today}

\begin{abstract}
A haloscope of the QUAX--$a\gamma$ experiment, composed of an  high-Q resonant cavity immersed in a 8 T magnet and cooled to $\sim 4.5$~K is operated to search for galactic axion with mass $m_a\simeq42.8~\mu\text{eV}$. The design of the cavity with hollow dielectric cylinders concentrically inserted in a OFHC Cu cavity,  allowed us to maintain a loaded quality-factor Q $\sim 300000$ during the measurements in presence of magnetic field. Through the cavity tuning mechanism it was possible to modulate the resonance frequency of the haloscope in the region $10.35337-10.35345$~GHz and thus acquire different dataset at different resonance frequencies. Acquiring each dataset for about 50 minutes, combining them  and correcting for the axion's signal estimation-efficiency we set a limit on the  axion-photon coupling $g_{a\gamma\gamma}< 0.731\times10^{-13}$ GeV$^{-1}$ with the confidence level set at $90\%$.

\end{abstract}


\maketitle


\section{\label{sec:intro}Introduction}
The axion is an hypothetical particle that was theorized to solve the strong CP problem. It arises from the spontaneous breaking of the Peccei-Quinn symmetry of QCD ~\cite{weinberg1978new,wilczek1978problem,peccei1977cp}. In addition, the properties predicted for the axion, charge neutrality, spin 0 and negligible interaction with the ordinary matter, make this particle a strong candidate for the dark matter~\cite{preskill1983cosmology}. Cosmology and astrophysical considerations, suggest an axion  mass range $1~\mu\text{eV} < m_a < 10~\text{meV}$~\cite{irastorza2018new}. The hunt for axion is now world spread and most of the experiments involved in this search use detectors based on the haloscope design proposed by Sikivie~\cite{sikivie1983experimental,PhysRevD.32.2988}.
Among them are ADMX~\cite{braine2020extended,du2018search,boutan2018piezoelectrically,bartram2021search}, HAYSTAC ~\cite{backes2021quantum,zhong2018results}, ORGAN~\cite{mcallister2017organ}, CAPP-8T~\cite{choi2021capp,lee2020axion}, CAPP-9T~\cite{jeong2020search}, CAPP-PACE~\cite{kwon2021first}, CAPP-18T ~\cite{lee2022searching}, GrAHal ~\cite{grenet2021grenoble}, RADES~\cite{melcon2020scalable,melcon2018axion,alvarez2021first}, TASEH~\cite{chang2022first}, QUAX~\cite{alesini2019galactic,alesini2021search,barbieri2017searching,crescini2018operation,crescini2020axion}, and KLASH~\cite{gatti2018klash,alesini2019klash}.
Dielectric and plasma haloscopes have also been proposed, like MADMAX~\cite{caldwell2017dielectric} and ALPHA ~\cite{lawson2019tunable}, respectively.
The haloscope concept is based on the immersion of a resonant cavity in a strong magnetic field in order to stimulate the inverse Primakoff effect, converting an axion into an observable photon~\cite{al2017design}. To maximise the power of the converted axion, it's necessary to maximise the cavity quality-factor ($Q$) and to tune the resonance frequency to match the axion mass.  Different solutions have being adopted to maximize the signal-to-noise ratio, facing the problem from different angles. Resonant cavities of superconductive and dielectric materials are becoming increasingly popular because of their high Q~\cite{di2019microwave,ahn2019maintaining,alesini2020high,alesini2021realization}.
In this work we describe the results obtained operating the haloscope of the  QUAX--$a\gamma$ experiment using an high-Q dielectric cavity immersed in a static magnetic field of 8 T and cooled down to $ \sim 4.5$  K.  The results obtained allow us to exclude values of $g_{a\gamma\gamma} > 0.729 \times10^{-13}$~GeV$^{-1}$ at 90\% confidence level (C.L.) in a region of mass 1.32 neV wide, centered at $42.8216\, \mu$eV.

\section{\label{sec:setup}Experimental Setup}

\subsubsection{General description}

The core of the haloscope is an extremely high-Q resonant cavity. The cavity is extensively described in \cite{PhysRevApplied.17.054013}: it is based on a right circular copper cavity with hollow sapphire cylinders that confine higher order modes around the cylinder axis.  The useful mode is a TM030, which has an effective volume $V\cdot C_{030}=3.4 \times 10^{-2}$ liters at the resonant frequency of 10.3 GHz, where $C_{030}$ is a geometrical factor entering in the signal-power estimation in Eq. \eqref{eq:power}. Under an 8\,T-field we measured an internal quality factor of more than $9\times 10^6$. The cavity and the magnet are hosted inside a liquid-He cryostat at a temperature of about 4 K.

\begin{figure}[htb]
  \centering
      \includegraphics[width=0.45\textwidth]{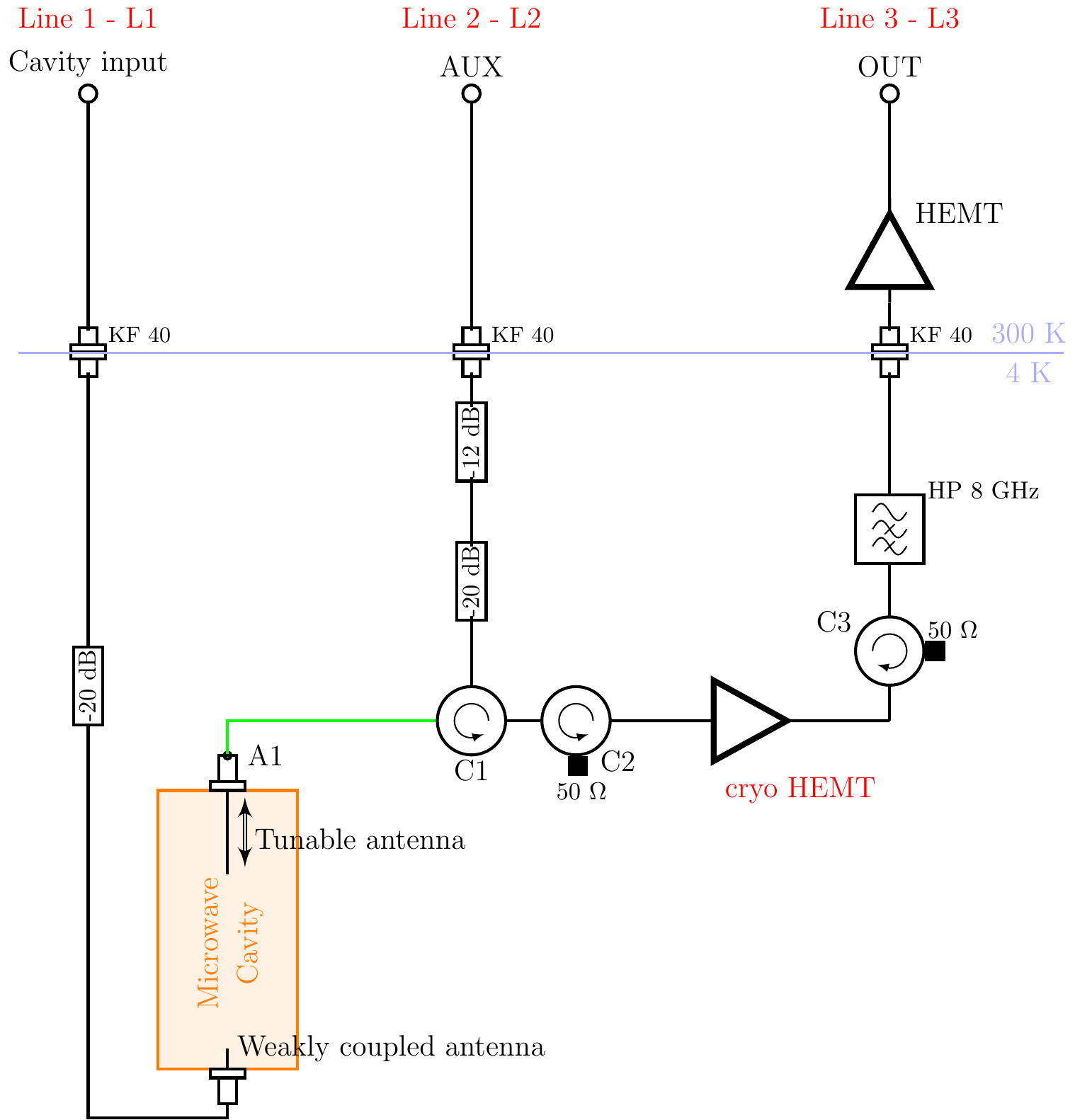}
\caption{\small Schematics of the experimental apparatus. The microwave cavity (orange) is immersed in the uniform magnetic field (not shown) generated by the magnet. C1, C2 and C3 are circulators, HP is a 8 GHz high pass filter, cryo HEMT and HEMT are High Electron Mobility Transistors. 
Attenuators are shown with their reduction factor in decibels. The horizontal blue line identifies the boundaries of the cryogenic stages of the apparatus. KF 40 are rf feed-through with ISO-KF 40 flanges.  }
\label{fig:Apparatus}
\end{figure}

The principle scheme of the measurement set up is shown in Figure \ref{fig:Apparatus}. The microwave cavity is immersed in a 8 T maximum magnetic field, not shown in the figure, generated by a 150~mm diameter bore of  500~mm length superconducting magnet. When the magnet is driven by a 92 A current, the effective squared field  over the cavity length amount to 50.8 T$^2$. 
The microwave cavity is read by a tunable monopole antenna with coupling $\beta$. This is obtained by acting on a manually-controlled mechanical feed-through, that allows for $\beta$ values in the range 0.01 to 20. A weakly coupled port (coupling about 0.01) is used for calibration purposes and is connected to the room temperature electronics by means of line L1. To avoid thermal power inputs from room temperature, a 20 dB attenuation is inserted on L1. Cavity tuning was obtained by displacing triplets of 2\,mm-diameter sapphire rods relative to the top and bottom cavity endcaps \cite{PhysRevApplied.17.054013}. Again, independent motion of the two triplets is obtained by manually controlled mechanical feed-throughs.

The power collected by the tunable antenna is amplified by a cryogenic high electron mobility transistor amplifier (cryo HEMT in figure), isolated from the cavity by means of the circulators C1 and C2. The output of the cryo HEMT is filtered and then transmitted along line L3 to the room temperature electronics, where it is first amplified by a room temperature HEMT and then processed for data storage. The room temperature chain is the same used in our previous measurements \cite{alesini2021search}: the HEMT output is frequency down-converted using a mixer with the local oscillator frequency set to a value about 500 kHz below the cavity resonance. The low frequency in phase and quadrature outputs of the mixer are amplified and then sampled with a 2 Ms/s analog to digital converter (ADC) and stored on a computer for off line data analysis.
Data storage is done with blocks of about 4 s of sampled data for both output channels of the mixer.
An auxiliary line L2 is used for calibration purposes: it is connected to the line L3 by means of the circulator C1, and 32 dB of attenuation prevents thermal leakage from room temperature components.

The room temperature electronic features also a Vector Network Analyser (VNA) for measurement of the scattering parameters S12 (input from line L2 - output from line L1), S31 and S32. From these scattering parameters it is possible to derive the loaded quality factor $Q_L$, resonance frequency $f_c$ and coupling $\beta$ of the tunable antenna. A diode noise source, having an equivalent noise temperature of about $10^4$ K,  can be fed to line L1 for testing after being amplified in such a way to have an equivalent noise temperature inside the microwave cavity slightly in excess of the thermodynamic temperature. A microwave signal generator and a microwave spectrum analyser are  used for the measurement of the  system noise temperature as described below. All rf generators, the VNA and the spectrum analyser are frequency locked to a GPS disciplined reference oscillator.

Following the figure, all components below the horizontal blue line sectioning the 4K region are enclosed in a vacuum chamber immersed in a liquid helium cryostat. A Ruthenium Oxide thermometer  measures the temperature of the cavity. 

\subsubsection{Data taking}\label{sec:datataking}

In a data taking run performed over a two weeks period in June 2021 we searched for an axion signal as a cavity excess-power in a small frequency band about 10.353 GHz, i.e. about an axion mass of 42.8 $\mu$eV. As shown in \cite{PhysRevApplied.17.054013}, the unloaded $Q$ factor of our microwave cavity is of several millions, well in excess of the axion quality factor  $Q_a = 10^6$.

We decided to perform measurements for three different values of the cavity loaded quality-factor $Q_L = Q_0/(1+\beta)$: 
\begin{enumerate}
\item[1)] $\beta \simeq 1$, i.e. $Q_L > Q_a$ 
\item[2)] $\beta \simeq 6$, i.e. $Q_L \simeq Q_a$ 
\item[3)] $\beta \geq 14$, i.e. $Q_L \ll Q_a$
\end{enumerate}

The total data taking session comprised 8 sub-runs in regime 1, 33 sub-runs in regime 2 and 11 sub-runs in regime 3. 
We performed the following steps for each sub-run:
\begin{enumerate}
\item[a.] looking to the S32 spectra with a VNA, we moved the cavity frequency to the desired value by acting (inserting) the sapphire triplet for tuning. Normally, for each sub-run a shift of half the cavity linewidth with respect to the previous sub-run was done. 
\item[b.] we stored the S32 spectra for the resulting cavity-configuration. This spectrum corresponds to a reflection-type measurement for the tunable port of the cavity.
\item[c.] we injected along the line L1 a white noise produced with the amplified noise-source and collected down-converted low-frequency I and Q spectra with the ADC. For this run, thermal-input spectra were usually integrated for about 2-3 minutes.
\item[d.] we removed any input to the system and collected data with the ADC. This step has been chosen to last 750 data blocks, 
for a total time of 3000 s. For each sub-run both the cavity temperatures and the liquid-helium level in the cryostat are 
recorded.
\end{enumerate}

The total time needed for a single sub-run is about one hour. We performed measurements only during the day, so that a typical day started with the liquid helium refilling of the cryostat, followed by the magnet charging lasting about 40 minutes. Considering 
that at the end of the day we needed to discharge the magnet for safety reasons, in a typical day we recorded about 5-6 sub-runs.

In this paper we present the results obtained for the sub-runs with strong cavity-coupling such that $Q_L \ll Q_a$ (regime 3), 
while the other data will be subject of a different paper. Indeed, extracting the axion signal in a regime where $Q_L \simeq Q_a$ 
or even higher poses a series of issues regarding systematics that necessitate a dedicated study. For the measurements in regime 
3 we dealt with a loaded quality-factor $Q_L \sim 3 \times 10^5$, which is much larger than the typical values used by other running-haloscopes, where such value was never in excess of $10^5$ (with the exemption of our previous measurement \cite{alesini2021search}). 

\subsubsection{Noise temperature and gain}

\begin{figure}[htb]
  \centering
      \includegraphics[width=0.4\textwidth]{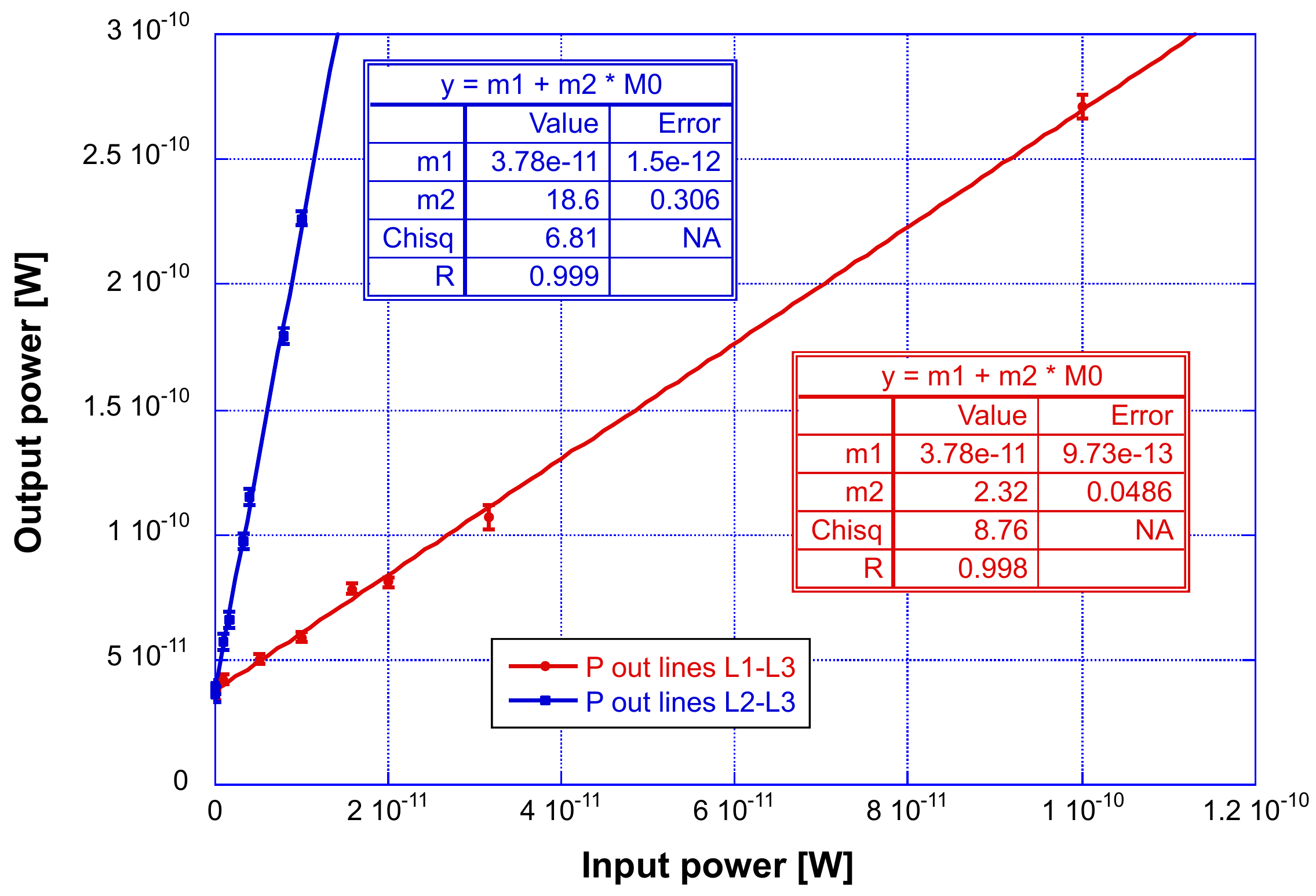}
\caption{\small Power output at the line L3 with variable input at the lines L1 (red) and L2 (blue). 
For the L1 input a rf signal at the cavity resonance-frequency is used, while for the L2 input the frequency is detuned by 1 MHz from the cavity resonance. Measurements are performed with a spectrum analyser taking 500 RMS averages of a 100 MHz window with a resolution bandwidth $B=1$ MHz.    }
\label{fig:Noiset}
\end{figure}

 We measured the system noise-temperature at the beginning, at the end and in the middle of the global data taking period. This procedure \cite{twpalnl} consists in measuring precisely the gains of the three lines L1, L2 and L3 from the point A1 in Figure \ref{fig:Apparatus}. The knowledge of the gains allows us to extract the system noise-temperature from the measurement of the noise level at the output of line L3.
 Gain measurement is obtained by feeding a calibrated power-level with the signal generator either from L1 or L2, and reading the outputs at L3 for inputs from both inputs lines or at L1, for an input from L2 only. The last measurement is only possible when 
 the tunable antenna has a significant coupling to the cavity, therefore only at the cavity resonant frequency. On the contrary, 
 to exploit the cavity reflection, the measurement from L2 to L3 is done at a frequency just off the cavity resonance. Figure \ref{fig:Noiset} shows the power measured at the output of line L3 by feeding power into L1 (red points) or L2 (blue points). 
 For each measure, we estimated the gains and the intercepts at zero, $P_0$, with a fit with a linear function. The transmission coefficient from input L2 to output L1 is done in a similar way. Combining these results, we computed the gains from A1 to L1, 
 L2, L3 as ($g_1$, $g_2$, $g_3$) = (-48.3, -39.3, 52) dB, respectively.
 We computed the system noise-temperature as
 \begin{equation}
     T_{\rm sys}= \frac{P_0}{k_B B g_3}
 \end{equation}
obtaining $T_{\rm sys} = 17.3 \pm 1$ K. Here, $k_B$ is the Boltzmann constant and $B$ the resolution bandwidth. 
This particular value of the noise temperature was obtained with the magnetic field on at the end of the last day of run, at the end of the sessions with regime~3. Previous measurements, performed with the magnet off, were in agreement with this one. The cause of the large noise-temperature observed was identified in a malfunction of the cryo HEMT exhibiting a quite high added noise. After the run, we measured its added noise separately finding a value about 10-12 K. 
 
\subsubsection{Raw data processing}

As described in \ref{sec:datataking}, we measured, for each subrun, the values of the cavity parameters by taking a reflection spectrum on the pick-up antenna, and a transmission spectrum with a thermal-noise source feeding power into the weakly coupled antenna. Parameters are extracted by fitting the spectra. A standard Lorentzian line shape is used to fit the transmission spectrum:

\begin{equation}
    S31^2(\nu)=\frac{A}{2 \pi}\frac{\Gamma}{(\nu-\nu_c)^2+(\Gamma/2)^2}
\end{equation}

With this equation we extract from the fit the cavity resonance frequency $\nu_c$ and the linewidth $\Gamma$, directly related to the loaded factor of merit $Q_L = \nu_c/\Gamma$. $A$ is a normalization constant.

A modified reflection function is used for the reflection spectrum to take into account some impedance mismatch between the cavity and the first-stage amplifier:
\begin{equation}
\label{s11}
    S32(\delta) = C\,\Bigg| \frac{\beta - 1 - i Q_0 \delta}{\beta + 1 + i Q_0 \delta} + i c\Bigg|
\end{equation}
where $C$ is a normalization constant, $\delta = \nu/\nu_c-\nu_c/\nu$, with $\nu_c$ the cavity resonance frequency, $\nu$ the frequency, $c$ is a free parameter related to the impedance mismatch, $Q_0$ is the cavity unloaded quality factor.
From this second fit we obtain the value of the coupling $\beta$.

\begin{figure}[h!]
  \centering
      \includegraphics[width=0.47\textwidth]{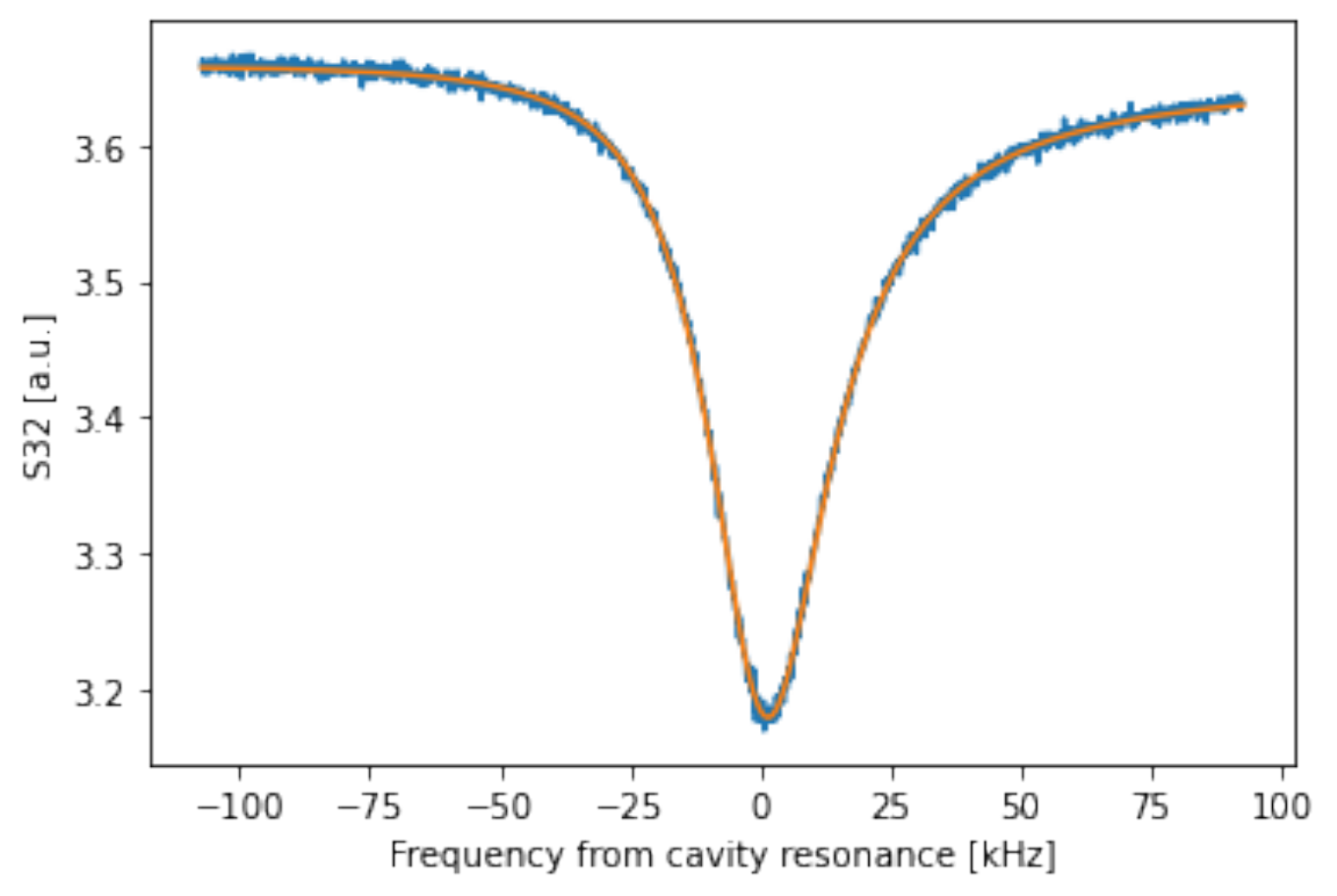}
\caption{\small Reflection spectrum obtained with the VNA and fit with the function (\ref{s11}). The fit results in the following values: $\beta = 14.59 \pm 0.01$, $\nu_c = 10 353 366 689 \pm 20$ Hz, $Q_0 = 5 565 000 \pm 8000$, $c = 0.0127 \pm 0.0001$, $C=3.6526\pm0.0002$. }
\label{fig:S11}
\end{figure}

\section{\label{sec:results}Data Analysis and Results}

By tuning the cavity resonance frequency ($\nu_c$) we acquired 11 different dataset, one for each $\nu_c$ in the range $10.35337-10.35345$~GHz (Tab.~\ref{tab:dataset}). For each dataset we calculated a power spectrum.
\begin{table}[htb]
  \begin{center}
    \caption{cavity resonance frequency, quality factor and cavity-antenna coupling for each dataset.}
    \label{tab:dataset}
  \vspace*{0.5cm}
    \begin{tabular}{c|c|c}
			\hline\hline
            $\nu_c$ [GHz]& $Q_L$ &  $\beta$   \\\hline
            10.3533667 & 365730 & 14.59  \\
			10.3533711 & 337630 & 15.91\\
			10.3533792 & 315100 & 17.00\\
			10.3533874 & 288190 & 18.00\\
			10.3533955 & 286620 & 17.87 \\
			10.3534036 & 284810 & 17.66\\
			10.3534159 & 283410 & 17.61 \\
            10.3534150 & 354000 & 13.74\\
			10.3534250 & 292510 & 16.20 \\
			10.3534354 & 290290 & 16.42\\
			10.3534464 & 285760 & 17.25 \\
			\hline
			\hline
    \end{tabular}
  \end{center}
\end{table}
In this section, we discuss the cumulative results obtained from the combined spectra. For sake of simplicity, examples from a single dataset are reported when necessary. 
The expected power generated by the axion conversion inside the haloscope is given by~\cite{BrubakePRL,al2017design}:
\begin{equation}\begin{split}
	\label{eq:power}
	P_{a}=\left( \frac{g_{a\gamma\gamma}^2}{m_a^2}\, \hbar^3 c^3\rho_a \right)\,
	\left( \frac{\beta}{1+\beta} \omega_c \frac{1}{\mu_0} B_0^2 V C_{030} Q_L \right)& \\
    \times \left(\frac{1}{1+\left(2Q_L \, \Delta_\omega  / \omega_c \right)^2}\right)
	\end{split}
    \end{equation}

\begin{figure}[h!]
  \centering
      \includegraphics[width=0.47\textwidth]{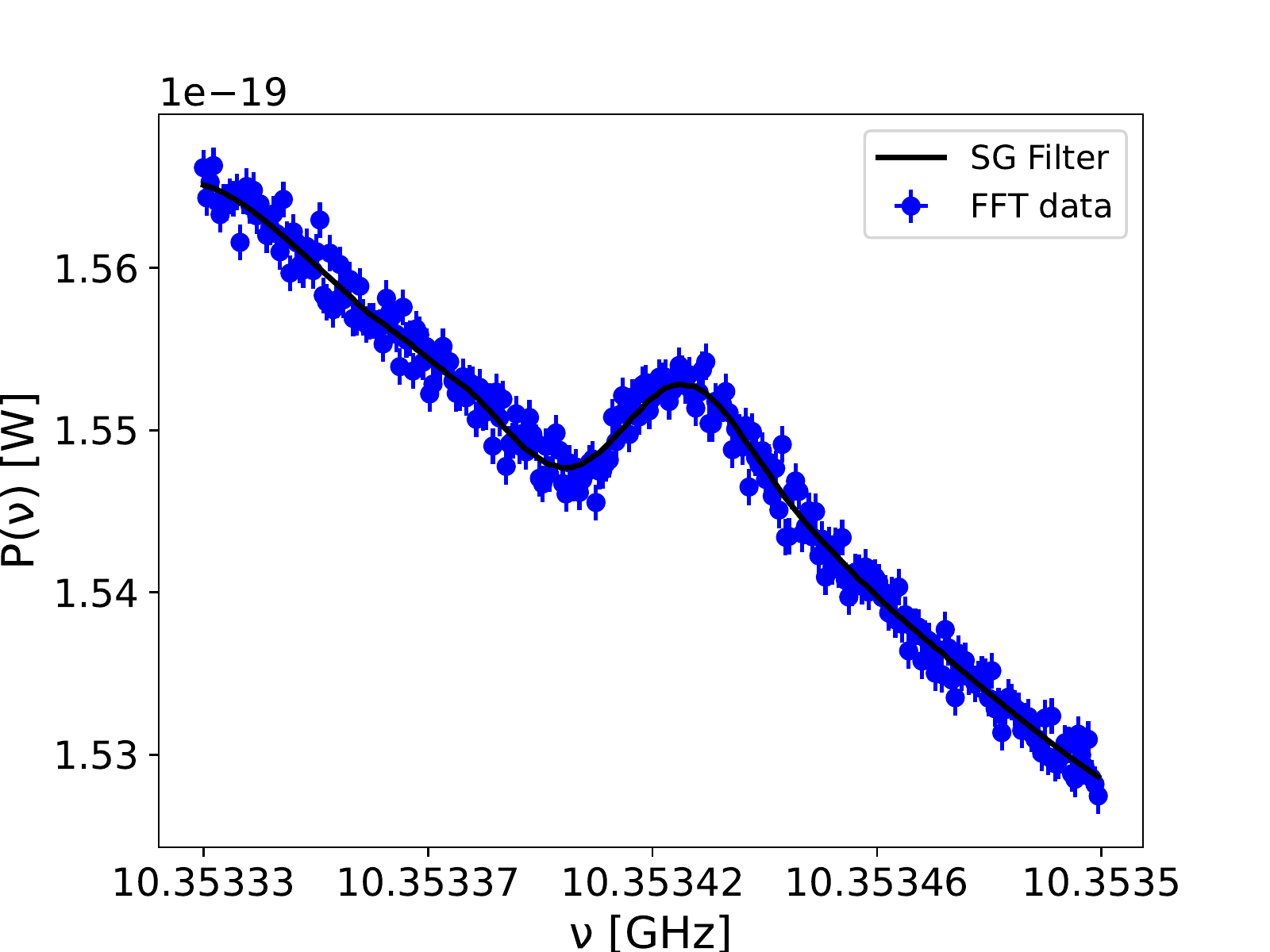}
\caption{\small FFT cavity power spectrum (blue dots) and SG filter (black line). $\nu_c=10.3534149$ GHz, $Q_L=354000$}
\label{fig:SG+data}
\end{figure}

\begin{figure}[h!]
  \centering
     \includegraphics[width=0.47\textwidth]{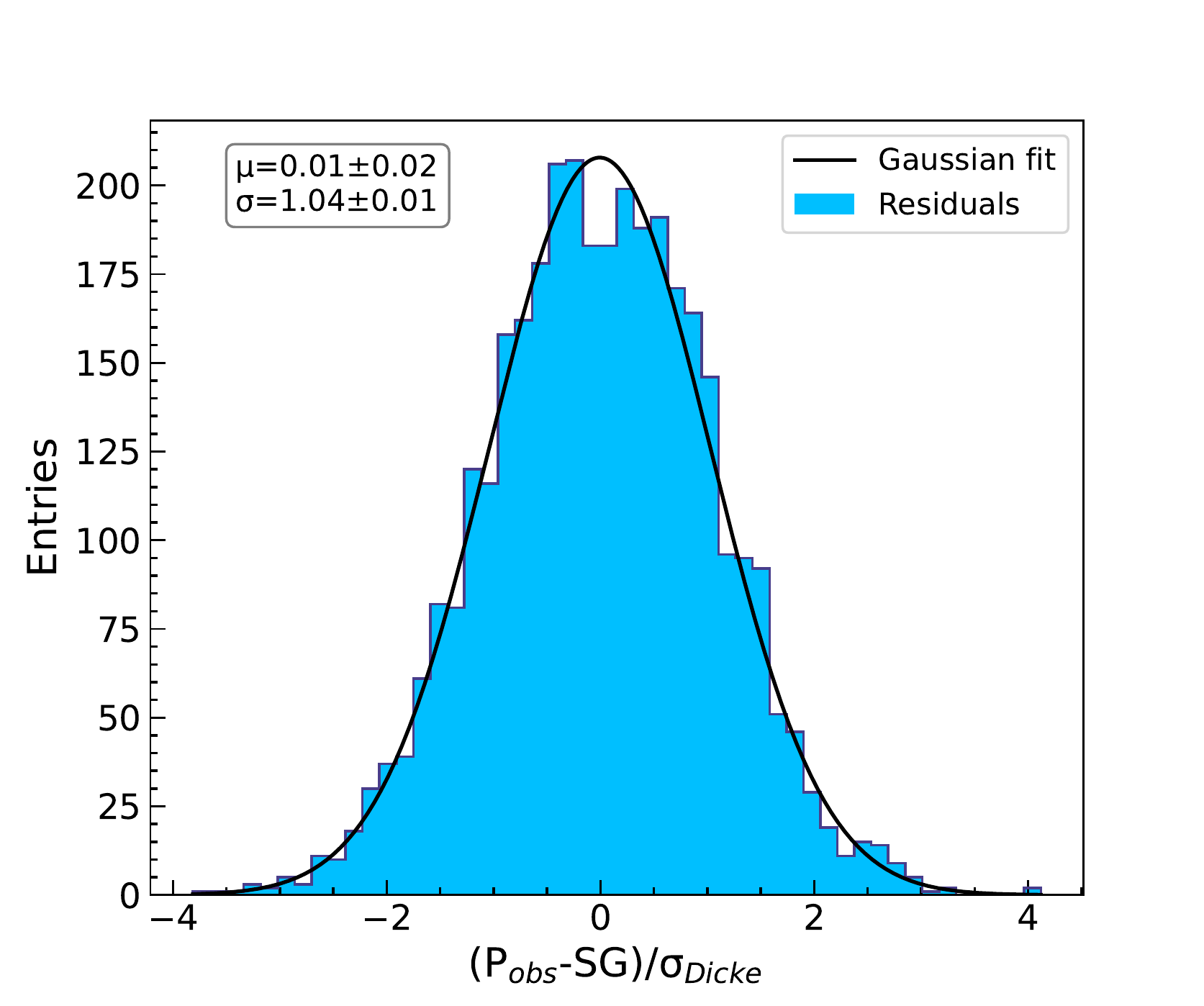}
\caption{\small Distribution of the cumulative residuals from each dataset normalized to the 
$\sigma_{\scriptscriptstyle \textup{Dicke}}$.}
\label{fig:cumulative_residual}
\end{figure}

In the first set of parenthesis, $\rho_a \sim  0.45$\,GeV/cm$^3$  ~\cite{10.1093/ptep/ptaa104} is the local dark matter density, $g_{a\gamma\gamma}$ is the coupling constant of the axion-photon interaction, $m_{a}$ is the axion mass.
The second set of parenthesis contains the vacuum permeability $\mu_0$, the magnetic field $B_0$ and the volume $V$ of the cavity. $\omega_c=2\pi\nu_c$ is the resonance angular frequency of the cavity, 
$\beta$ and $Q_L$ are antenna coupling and loaded quality factor as described above.
$C_{030}$ is a geometrical factor equal to about 0.028 for the TM030 mode of this cylindrical dielectric cavity. 
In the third brackets, a Lorentzian function describes the effect of the detuning $\Delta_w = \omega_c-\omega_a$ between the cavity and an axion having angular frequency $\omega_a$.
In presence of a signal due to axion conversion a power excess would be observable in the residuals of the power spectrum. The residuals are obtained subtracting a Savitzky-Golay (SG) filter~\cite{savitzky1964smoothing} of the fourth order to the cavity power spectrum. The dynamic interval of the SG filter was optimized to 59.2 kHz (91 bins). For each dataset, we applied the SG filter to a window [$\nu_c-3\Gamma, \nu_c+3\Gamma$] of about 180 kHz, corresponding to six linewidths $\Gamma$ and centered on the cavity resonance frequency $\nu_c$ as showed in Figure \ref{fig:SG+data}.
In the laboratory frame, the axion signal is expected to have a width of about 10 KHz ~\cite{sikivie1983experimental, turner1990periodic}. With a power spectrum with bin width $\Delta \nu=651$ Hz we expect the axion signal to be distributed over 16 consecutive bins.
We normalized the residuals of each dataset to the expected noise power $\sigma_{\scriptscriptstyle \textup{Dicke}}$ calculated 
with the Dicke radiometer equation ~\cite{dicke1946measurement}
\begin{equation}
  \sigma_{\scriptscriptstyle \textup{Dicke}} = k_B T_{\rm sys} \sqrt{\Delta \nu/\Delta t}\, ,
\end{equation}
where $T_{\rm sys}$, is the system noise-temperature,
$\Delta \nu$ is the bin width (651 Hz) and $\Delta t$ is the integration time (3000 s). The distribution of the cumulative normalized-residuals from all the datasets is shown in Fig.~\ref{fig:cumulative_residual} along with a Gaussian fit, showing a standard deviation compatible with 1.

We use the Least-Squares method to estimate the best value $\hat{g}_{a\gamma\gamma}$ for the axion-photon coupling, by 
minimizing    
\begin{equation}
 \chi^2 = \sum_{\alpha=1}^{N_{\text{scan}}} \sum_{i=1}^{N_{\text{bin}}}\,
\left[\frac{R^{(\alpha)}_i - S^{(\alpha)}_i (m_a, g_{a\gamma\gamma}^2)}{\sigma^{(\alpha)}_{\text{Dicke}}}\right]^2\,,
\label{eq:chisquare}
\end{equation}
where the $\alpha$ index runs over the $N_{\text{scan}}$ datasets taken with different cavity resonant-frequencies, the index $i$ runs over the frequency bins of each power spectrum, $R_{i\alpha}$ and $S_{i\alpha}$ are the residuals and the expected power 
signals for the scan $\alpha$ and  frequency bin $i$, respectively. $S_{i\alpha}$ is calculated as the integral in the frequency domain of equation (\ref{eq:power}) multiplied by the spectrum of the full standard halo model distribution \cite{turner1990periodic}.

We express the expected power as $S_{\alpha,i} (m_{a},g_{a\gamma\gamma}^2) = g_{a\gamma\gamma}^2\,T_{\alpha,i} (m_{a})$, and analytically minimize Eq. (\ref{eq:chisquare}) by solving $\partial \chi^2 / \partial g^2_{a\gamma\gamma} = 0$, and 
calculating the uncertainty according to the formula $(\xi = g^2_{a\gamma\gamma})$:
\begin{equation}
\frac1{\sigma^2_{\hat{\xi}}} = \frac12\,\frac{\partial^2 \chi^2}{\partial \xi^2}\,.
\end{equation}

Solving this equation, we get: ($\sum\sum \equiv \sum_{\alpha=1}^{N_{\text{scan}}} \sum_{i=1}^{N_{\text{bin}}}$)
\begin{equation}
%
%
  \overline{g^2} =  \sigma^2 (\overline{g^2})\,\left[\sum\sum\,\frac{R^{(\alpha)}_i\,T^{(\alpha)}_i (m_a)}{(\sigma^{(\alpha)}_{\text{Dicke}})^2}\right]\,
\label{eq:G2}
\end{equation}
where $\overline{g^2}$ is the average squared coupling-constant that accounts for the contributions of all the frequency bins 
of all the datasets, and
\begin{equation}
\sigma^2 (\overline{g^2}) = \sum\sum \, \left[\frac{T^{(\alpha)}_i (m_a)}{\sigma^{(\alpha)}_{\text{Dicke}}}\right]^2
\label{eq:sigmaG2}
\end{equation}
is its variance. 
We repeated this procedure for different values of $m_a$ and calculated $\overline{g^2}$ and $\sigma(\overline{g^2})$ 
for axions masses in the range $42.8210-42.8223$ $\mu$eV. 

\begin{figure}[h!]
  \centering
      \includegraphics[width=0.47\textwidth]{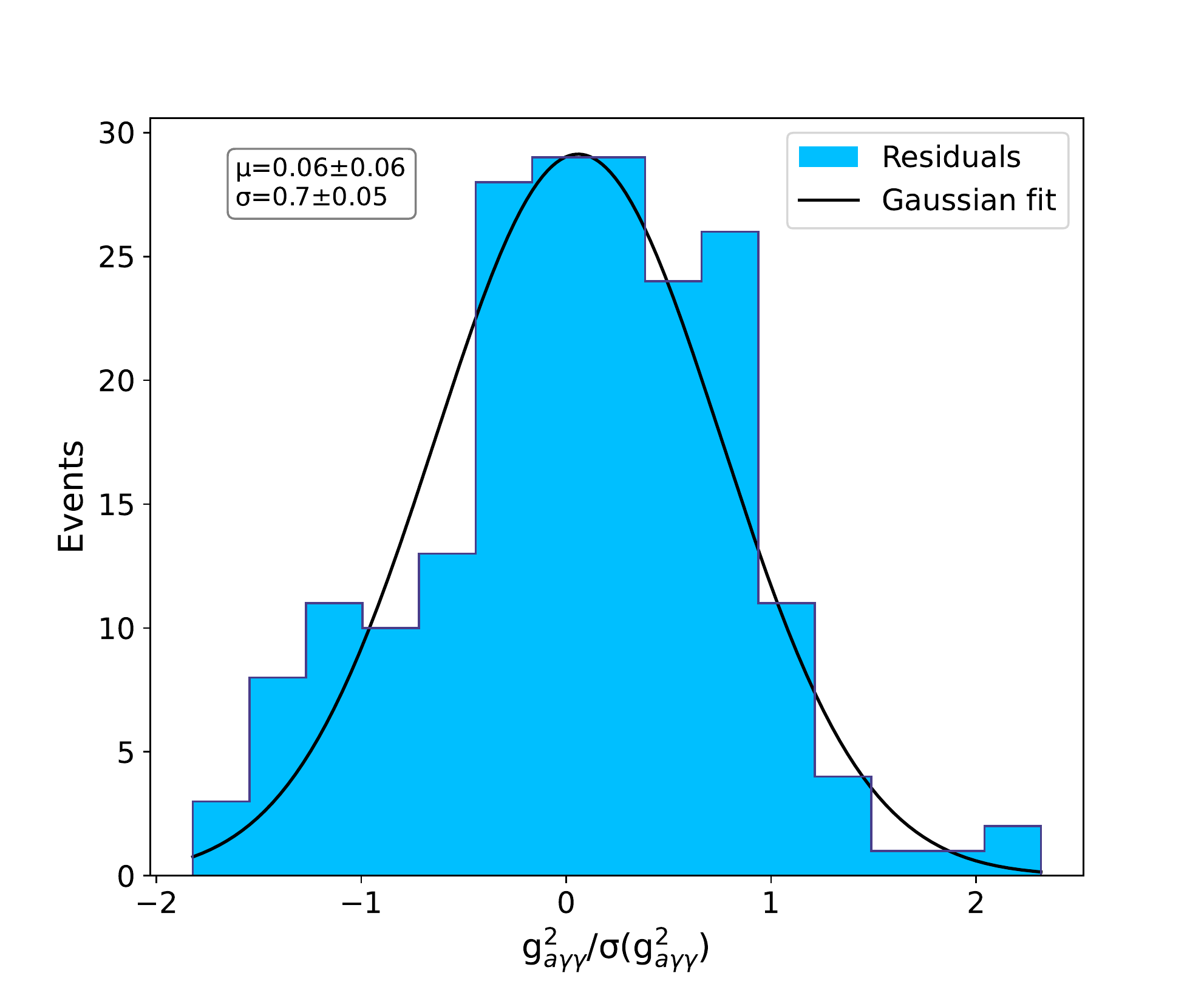}
\caption{\small Histogram of the $\overline{g^2}$/$\sigma(\overline{g^2})$ distribution calculated using Eqs. \eqref{eq:G2} and \eqref{eq:sigmaG2}. No excess above $5 \sigma$ was observed.} .
\label{fig:histog2}
\end{figure}

A candidate discovery requires the detection of a power excess larger than 5$\sigma$ above the noise, hence in the distribution of $\overline{g^2}$/$\sigma(\overline{g^2})$. We did not find any candidate (see Fig. \ref{fig:histog2}) and the result is interpreted as an exclusion test in this axion-mass range.

\begin{figure}[h!]
  \centering
      \includegraphics[width=0.47\textwidth]{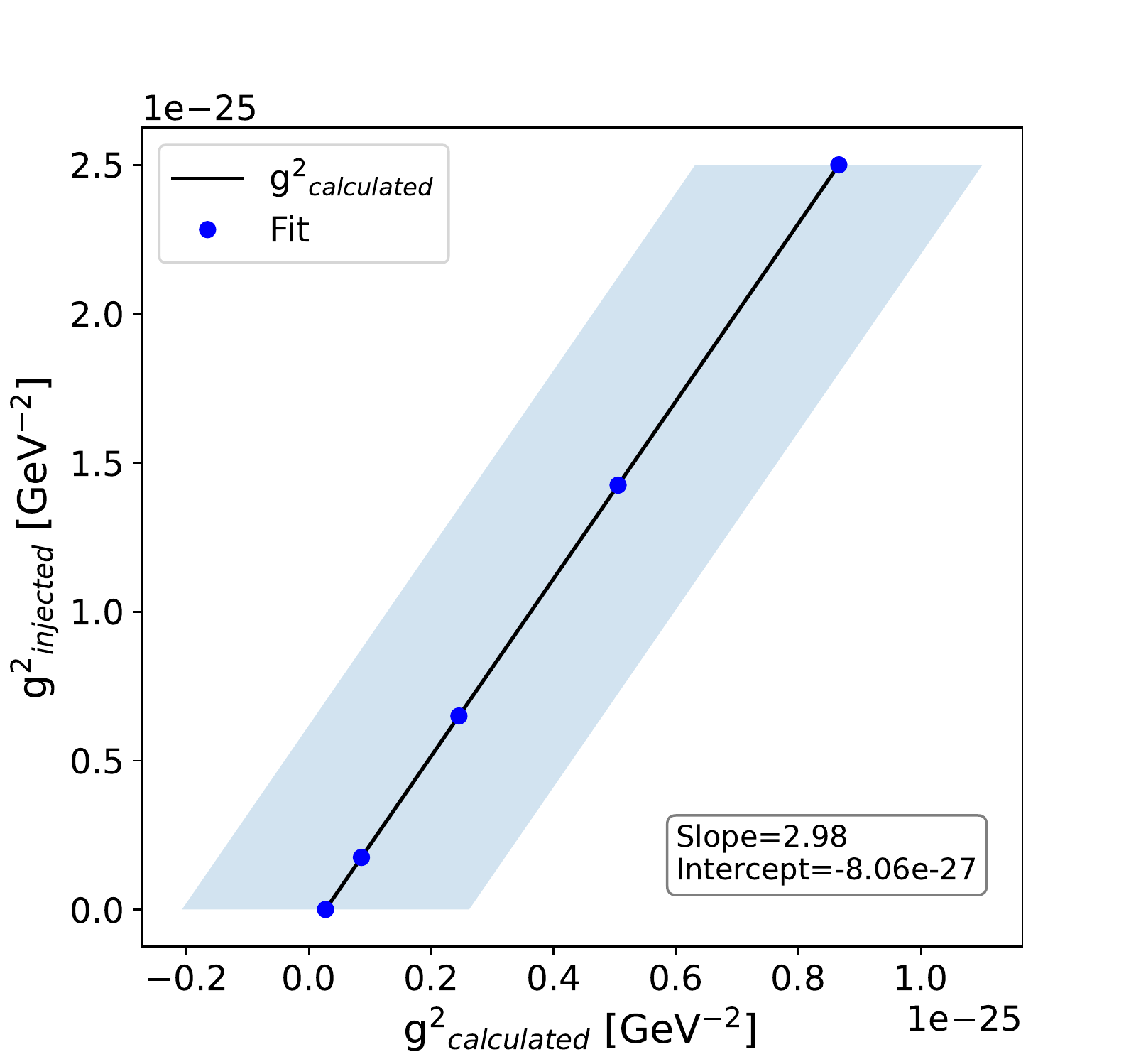}
\caption{\small Relation between g$_{\text{injected}}^2$ and the mean of the $g_{\text{calculated}}^2$ distribution (blue points), along with the best linear fit parameters. The belt represents the standard deviation of the distribution of $g_{\text{calculated}}^2$ obtained after 100 simulations.}
\label{fig:g2vsg2}
\end{figure}

So far, we did not consider the efficiency of the SG filter in estimating the axion signal. In order to quantify it, we run a 
Monte Carlo simulation where a fake axion signal, with a known $g_{\text{injected}}^2$, is numerically inserted in simulated 
power-spectra with different $\nu_c$. We used Eq. \eqref{eq:G2} to estimate the $g^2$ for each injected signal (g$g_{\text{calculated}}^2$), and determined the efficiency from the relation between $g_{\text{calculated}}^2$ 
and $g_{\text{injected}}^2$.
We simulated the cavity power-spectra by adding random Gaussian noise (mean=0, sigma=$\sigma_{Dicke}$, random values extracted between 0 and $\pm\,\sigma_{\text{Dicke}}$, according to a Gaussian PDF) to the SG filters. 
For a given axion mass, $m_a$, the estimation of the efficiency works as follows: 1) for each data set we calculate a simulated spectrum; 2) a fake axion signal with a known $g_{\text{injected}}^2$ is injected in the simulated spectra; 3) Eq. \eqref{eq:G2} is used to compute g$_{calculated}^2$; 4) points 2 and 3 are repeated for different values of $g_{\text{injected}}^2$; 5) points 2-4 are repeated for a new set of simulated spectra in order to increase the statistics. 
The output of this procedure is a distribution of $g_{\text{calculated}}^2$ for each value of $g_{\text{injected}}^2$. The relation between $g_{\text{injected}}^2$ and the mean of $g_{\text{calculated}}^2$, calculated accounting for the contribution of all the datasets, is shown in Figure \ref{fig:g2vsg2} for axion proper frequency f$_{axion}$=10.35341562 GHz, 
$g_{\text{injected}}^2 = [0.01,1.755,6.502,14.25,25] \times 10^{-26}$ GeV$^{-2}$, and for 100 simulated spectra.

The distribution of $g_{\text{injected}}^2$ {\it vs} $g_{\text{calculated}}^2$  shows a linear relation with slope very close to 3 and an intercept different from zero. These features are valid for all the axion masses not in the immediate proximity of the edge of 
the power spectrum, where the slope deviates considerably from 3. The slope of this linear relation is interpreted as the inverse 
of the estimation efficiency. If we neglect the intercept, then $g_{\text{calculated}}^2$/$g_{\text{injected}}^2 \sim 1/3$,  i.e. an estimation efficiency of about 0.33. The intercept represents a contribution to the $g_{\text{calculated}}^2$  given by the 
average noise of the simulated spectra for a given m$_a$.

Once we corrected the estimated $\overline{g^2}$ by the filter efficiency, we calculated the limit on the axion-photon coupling with a 90\% confidence level as in~\cite{alesini2021search}, using a power constrained procedure for the $\overline{g^2}$ that under fluctuates below $-\sigma$ ~\cite{cowan2011power}. 
\begin{figure}[h!]
  \centering
      \includegraphics[width=0.47\textwidth]{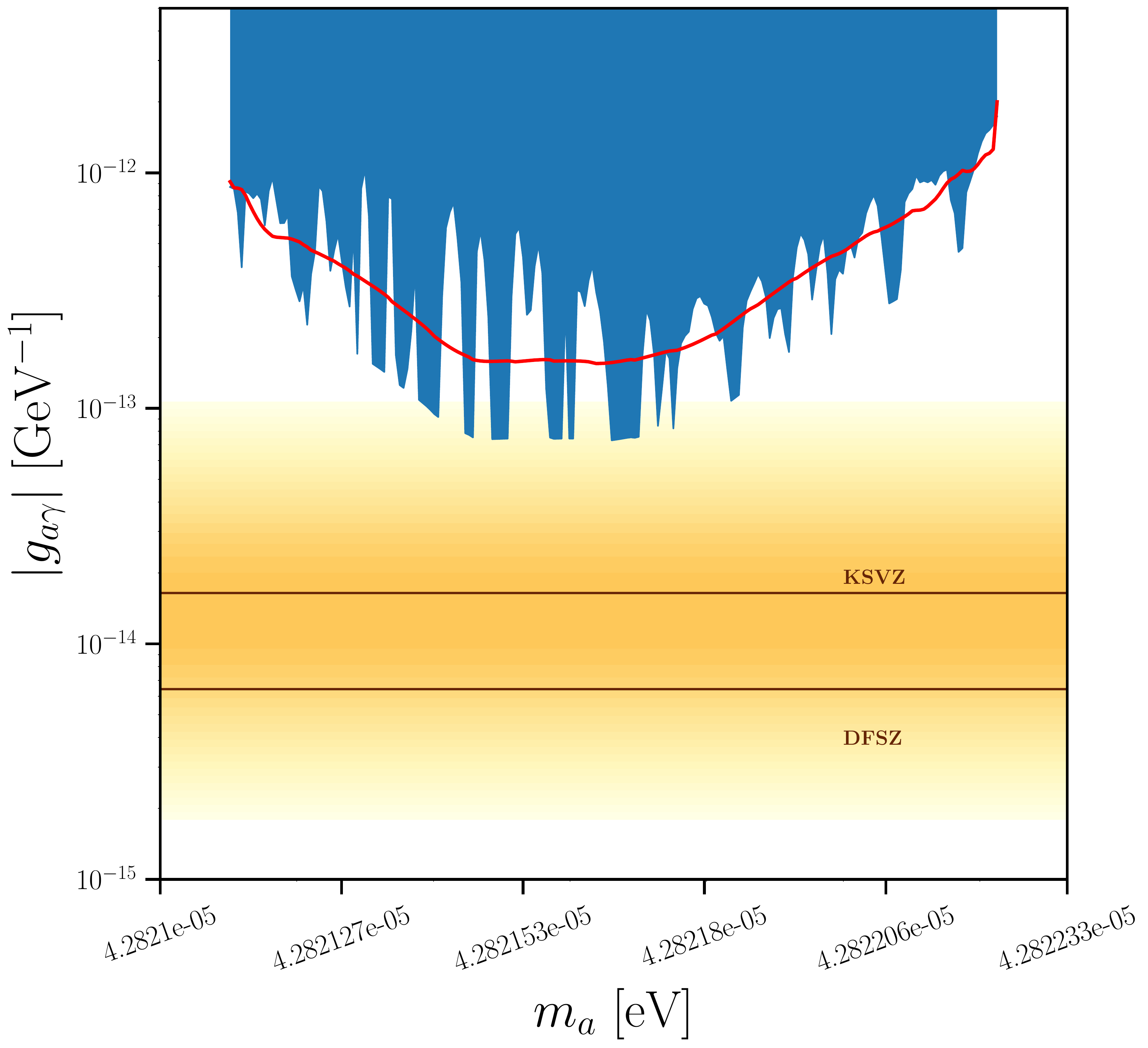}
\caption{\small The 90\% single-sided C.L. upper limit for the axion coupling constant $g_{a \gamma \gamma}$ as a function of the axion mass. The red solid curve represents the expected limit in the case of no signal. The yellow region indicates the QCD axion model band. Image realized using: https://github.com/cajohare/AxionLimits}
\label{fig:gCL}
\end{figure}
In Fig.~\ref{fig:gCL} we show the calculated upper-limit  $g_{a\gamma\gamma}^{\scriptscriptstyle \textup{CL}}$ in the axion mass range $42.821-42.8223$ $\mu$eV, i.e. a mass window of  about 1.32 neV centered in 42.8216 $\mu$eV. The reference upper-limit of this analysis is the value at the maximum sensitivity (the minimum spectrum reported in Fig.~\ref{fig:gCL}, $g_{a\gamma\gamma}^{\scriptscriptstyle \textup{CL}} < 0.731\times10^{-13}$~GeV$^{-1}$ at 90\% C.L.

\section{\label{sec:conclusions}Conclusions}
We reported the results of the search of galactic axions using an high-Q dielectric haloscope.  The investigated mass range is  $42.8210-42.8223$ $\mu$eV. We set a limit for the axion-photon coupling a factor about 4 from the axion-QCD band. We demonstrated the robustness of this detection approach and the importance of working with a high Q cavity in the search for axions. In fact, we managed to reach a sensitivity that almost touches the QCD band, even though the equivalent thermal noise of our system was very high (about 17 K) due to an experimental setback, and the detection efficiency was considered. In future experiments of this kind the sensitivity could be further improved, reducing the overall noise and improving the thermalization of the cavity.

\section*{acknowledgments}

DA VERIFICARE

We are grateful to E. Berto, A. Benato, and M. Rebeschini for the mechanical work; F. Calaon and M. Tessaro for help with the electronics and cryogenics. We thank G. Galet and L. Castellani for the development of the magnet power supply, and M. Zago who realized the technical drawings of the system. We deeply acknowledge the Cryogenic Service of the Laboratori Nazionali di Legnaro for providing us with large quantities of liquid helium on demand.
This work was supported by INFN  and partially supported by EU through FET Open SUPERGALAX project, Grant N.863313


\bibliography{biblioFile}


\end{document}